\documentclass{iopart}
\usepackage{iopams}
\usepackage{amssymb,amstext}
\usepackage{graphicx}

\begin{document}

\title[MD simulation of amorphous silica under high pressure]
{Molecular dynamics computer simulation of amorphous silica under high pressure}
\author{J~Horbach}
\address{Institut f\"ur Materialphysik im Weltraum, Deutsches Zentrum 
f\"ur Luft- und Raumfahrt (DLR), 51170 K\"oln, Germany}
\eads{\mailto{juergen.horbach@dlr.de}}


%
\begin{abstract} 
The structural and dynamic properties of silica melts under high pressure
are studied using molecular dynamics (MD) computer simulation. The
interactions between the ions are modeled by a pairwise-additive
potential, the so-called CHIK potential, that has been recently
proposed by Carr\'e {\it et al}.  The experimental equation of state is
well-reproduced by the CHIK model.  With increasing pressure (density),
the structure changes from a tetrahedral network to a network containing
a high number of five- and six-fold Si-O coordination.  In the partial
static structure factors, this change of the structure with increasing
density is reflected by a shift of the first sharp diffraction peak
towards higher wavenumbers $q$, eventually merging with the main peak
at densities around 4.2\,g/cm$^3$. The self-diffusion constants as a
function of pressure show the experimentally-known maximum, occurring
around a pressure of about 20\,GPa.
\end{abstract} 


%
\section{Introduction}
Under ambient conditions, amorphous silica (SiO$_2$) forms a disordered
tetrahedral network. Tetrahedral units with silicon atoms in the
centre are connected to each other via the oxygen atoms at the four
corners of each tetrahedron. Under compression, the network structure
of SiO$_2$ may change significantly.  From experiments on silica glasses
\cite{grimsditch84,hemley86,williams88,meade92,zha94,brazhkin03,lin07},
it is known that the SiO coordination changes for pressures of the order
of 15--20\,GPa.  Then, the network structure is dominated by five- and
six-fold coordinated SiO units.  Albeit of great interest for geoscience,
experimental studies of pure silica melts under (high) pressure are very rare
\cite{brazhkin03}. This is due to the high temperatures that are required
to study silica in its liquid state. Hence, computer simulations are an
important tool to elucidate structural and dynamic properties of silica
melts at high pressure.

In many of the recent molecular dynamics (MD) computer simulations,
the pair potential proposed by van Beest, Kramer and van Santen,
the so-called BKS potential \cite{bks90}, has been used to model
the interactions between the Si and O ions.  Tse {\it et al}
\cite{tse92} showed that in BKS silica the oxygen coordination
around silicon atoms changes from 4 to 5 at intermediate pressures
(10-15\,GPa) and reaches 6 at high pressures. Subsequent studies
of BKS silica \cite{barrat97,hemley00,horbach96} have indicated
that transport processes first become faster with increasing
pressure, in agreement with experimental studies on silicate melts
\cite{kushiro78,rubie93,poe97,tinker03}.  In Ref.~\cite{barrat97}, this
was interpreted as a signature of a strong-to-fragile transition,
according to Angell's classification of glassforming liquids
\cite{angell85,glassbook}.  Note that the anomalous transport in silica
under pressure was already seen in a pioneering MD simulation by Angell
{\it et al} \cite{angell92} using a different model potential. Further
studies on BKS silica \cite{saika00,shell02,sharma06,legrand07} addressed
the question to what extent the properties of silica are typical for a
strong glassformer, sharing its properties with other network-forming
liquids such as water. In Ref.~\cite{saika00}, the possibility of a phase
separation between a low and high density liquid phase was discussed
and a rough estimate of the free energy as a function of volume and
temperature was used to locate such a transition for BKS silica, finding
a critical point at a density of about 2.9\,g/cm$^3$ and a temperature
of about 2000\,K.

But recent works have called in question that simulations with
simple pair potentials can give reliable results for pressure-induced
phenomena in silica.  Trave {\it et al} \cite{trave02} claimed that
empirical force fields such as the BKS model ``are not sufficiently
accurate to reproduce pressure-induced phenomena such as changes in
coordination and increasingly strained geometries''. These authors used
``first-principles'' Car-Parrinello molecular dynamics (CPMD) simulations
to investigate liquid silica for pressures up to about 30\,GPa at a
constant temperature of $T=3500$\,K.  In a CPMD, the electronic degrees of
freedom are taken into account in the framework of a density functional
theory \cite{kohn96} and thus it is not relying on the use of effective
interatomic potentials.  Whereas the CPMD calculation for silica shows
good agreement with respect to the experimental equation of state, the
BKS model underestimates the experimental specific volume systematically
by about 15\%. Based on CPMD calculations, Tangney and Scandolo (TS)
\cite{tangney02} have developed a new effective potential that nicely
reproduces the experimental equation of state of amorphous silica as
well as some of the properties of crystalline phases \cite{herzbach05}.
The TS model is a fluctuating dipole moment potential, accounting for
the polarizability of the oxygen atoms. A disadvantage of the TS model
is its low efficiency. Due to the expensive self-consistent computations
of the dipole moments on the oxygen atoms in every time step, up to two
orders of magnitude more computer time is needed in MD simulations with
the TS model than for comparable simulations with a simple pair potential
such as the BKS model \cite{herzbach05}.

The conjecture of Trave {\it et al} about the validity of simple effective
interaction models for silica has been recently challenged by Carr\'e {\it
et al} \cite{carre07_1} who developed a new pair potential for silica, the
so-called CHIK potential.  The CHIK potential was also parametrized
from CPMD calculations and was shown to be superior to the BKS model
with respect to various properties of amorphous silica (in particular the
density at low pressures) \cite{carre07_1}.  In this work, we show that
it also reproduces very accurately the experimental equation of state,
various structural properties and the anomalous diffusion dynamics
of silica under pressure. This proves the transferability of the CHIK
potential to the description of amorphous silica under high pressure.
Thus, the CHIK model can be used to readdress the questions about the
possibility of a liquid-liquid phase separation and the mechanisms of
anomalous transport phenomena at high pressures.

\section{The model potential and details of the simulation}
\label{sec:phase}
At first glance, it might be surprising that the directional covalent
bonding in silica can be described by an interaction model that
depends only on the magnitude of the distance between atom pairs. In
fact, in a one-component system such as silicon one has to include
at least three-body interactions to reproduce a tetrahedral network
structure. However, in a binary system such as silica there is the
competition between three different pair potentials.  By tuning the
parameters for the repulsive OO interactions and the attractive SiO
interactions appropriately, one can generate a network structure with
Si-O tetrahedra as the preferred structural units.

\begin{figure}
\vspace*{0.26cm}
\begin{center}
\includegraphics[width=0.65\textwidth]{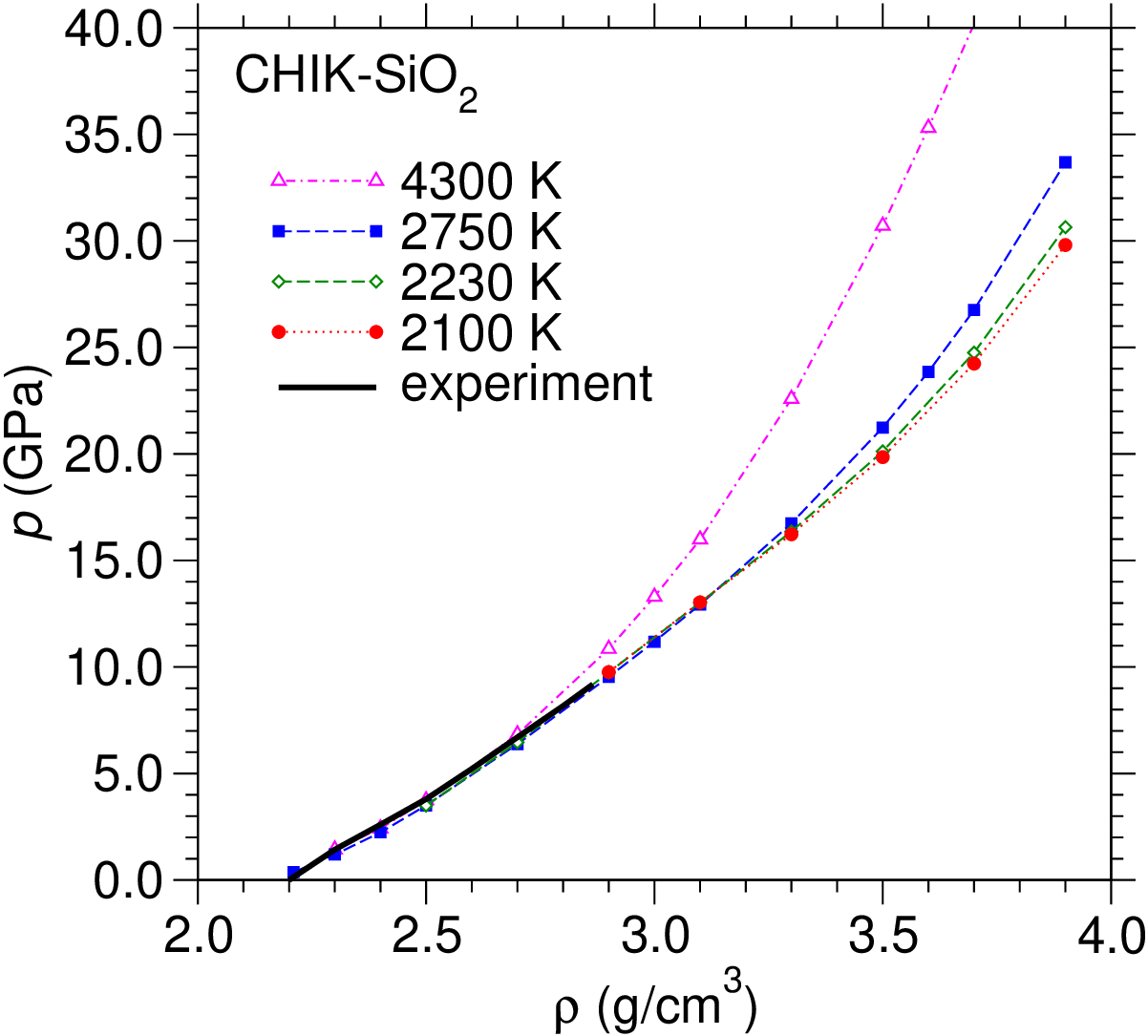}
\caption{Pressure as a function of density for several temperatures, 
as indicated. The experimental data are taken from Ref.~\cite{tsiok98}. 
\label{fig1} 
}
\end{center}
\end{figure}
Apart from a rather unreliable description of the equation state, the
widely used BKS potential \cite{bks90} has been proven successful to
describe various static and dynamic properties of amorphous silica quite
reliably (see, e.g., Ref.~\cite{vollmayr96,horbach99,horbach01}).  The
parametrization of the BKS model was based on Hartree-Fock calculations
of a single SiO$_4$ tetrahedron, charge-saturated by four hydrogen atoms
\cite{bks90}.  In contrast to that, the development of the CHIK potential
was based on a CPMD simulation of a bulk system of 114 particles. 
A structural fitting procedure was used for the parametrization.
The idea was to match the partial pair correlation functions, as obtained
with the new effective potential, with those obtained from CPMD. As the
BKS potential, the CHIK model consists of a long-ranged Coulomb part
and a short-ranged Buckingham potential,
\begin{equation}
  u_{\alpha\beta} (r) = \frac{q_{\alpha}q_{\beta}e^2}{r}
      + A_{\alpha \beta} \exp\left( - B_{\alpha\beta} r \right)
      - \frac{C_{\alpha\beta}}{r^6} ,
\end{equation}
where $r$ is the distance between an ion of type $\alpha$ and an
ion of type $\beta$ ($\alpha, \beta = {\rm Si, O}$), $e$ is the
elementary charge, and $q_{\rm Si}=1.910418$ and $q_{\rm O}=-q_{\rm
Si}/2$ are effective partial charges. The values of the parameters
$\{A_{\alpha\beta}, B_{\alpha\beta}, C_{\alpha\beta}\}$ are $A_{\rm
SiSi}=3150.462646$\,eV, $B_{\rm SiSi}=2.851451$\,\AA$^{-1}$, $C_{\rm
SiSi}=626.751953$\,eV/\AA$^6$, $A_{\rm SiO}=27029.419922$\,eV, $B_{\rm
SiO}=5.158606$\,\AA$^{-1}$, $C_{\rm SiO}=148.099091$\,eV/\AA$^6$,
$A_{\rm OO}=659.595398$\,eV, $B_{\rm OO}=2.590066$\,\AA$^{-1}$, and
$C_{\rm OO}=26.836679$\,eV/\AA$^6$. More details on the potential can
be found in Ref.~\cite{carre07_1}.

MD simulations have been performed for systems of $N=1152$ particles at
the densities $\rho=2.3$, 2.4, 2.5, 2.7, 2.9, 3.0, 3.1, 3.3, 3.5, 3.6,
3.7, 3.9, and 4.2\,g/cm$^3$ in the temperature range 6100\,K$\ge T \ge$
2100\,K.  The equations of motion were integrated by the velocity form
of the Verlet algorithm \cite{allen} using a time step of 1.6\,fs. The
long-range Coulomb forces and energies were computed by the Ewald
summation technique \cite{allen}.  At each state, the system was
equilibrated in the $NVT$ ensemble during 0.2\,ns to 25\,ns.  Temperature
was kept constant by coupling the system periodically to a stochastic
heat bath.  After the equilibration, structural and dynamic quantities
were obtained from production runs in the microcanonical ensemble.

Figure \ref{fig1} displays the equation of state, $p(\rho)$, along
different isotherms.  For pressures up to about 10\,GPa, the pressure is
almost independent of temperature in the range 4300\,K$\ge T \ge$ 2100\,K.
In this pressure range, the simulation is in very good agreement with
experiment \cite{tsiok98}.  This demonstrates the transferability of the
CHIK model: this model has been parametrized from a single CPMD run at
$T=3600$\,K and $\rho=2.2$\,g/cm$^3$, without considering states at high
pressure. Thus, at least as far as the equation of state is concerned,
the CHIK potential provides a reliable description and so it appears to
be appropriate for the investigation of silica under high pressure.

\begin{figure}
\vspace*{0.3cm}
\begin{center}
\includegraphics[width=0.65\textwidth]{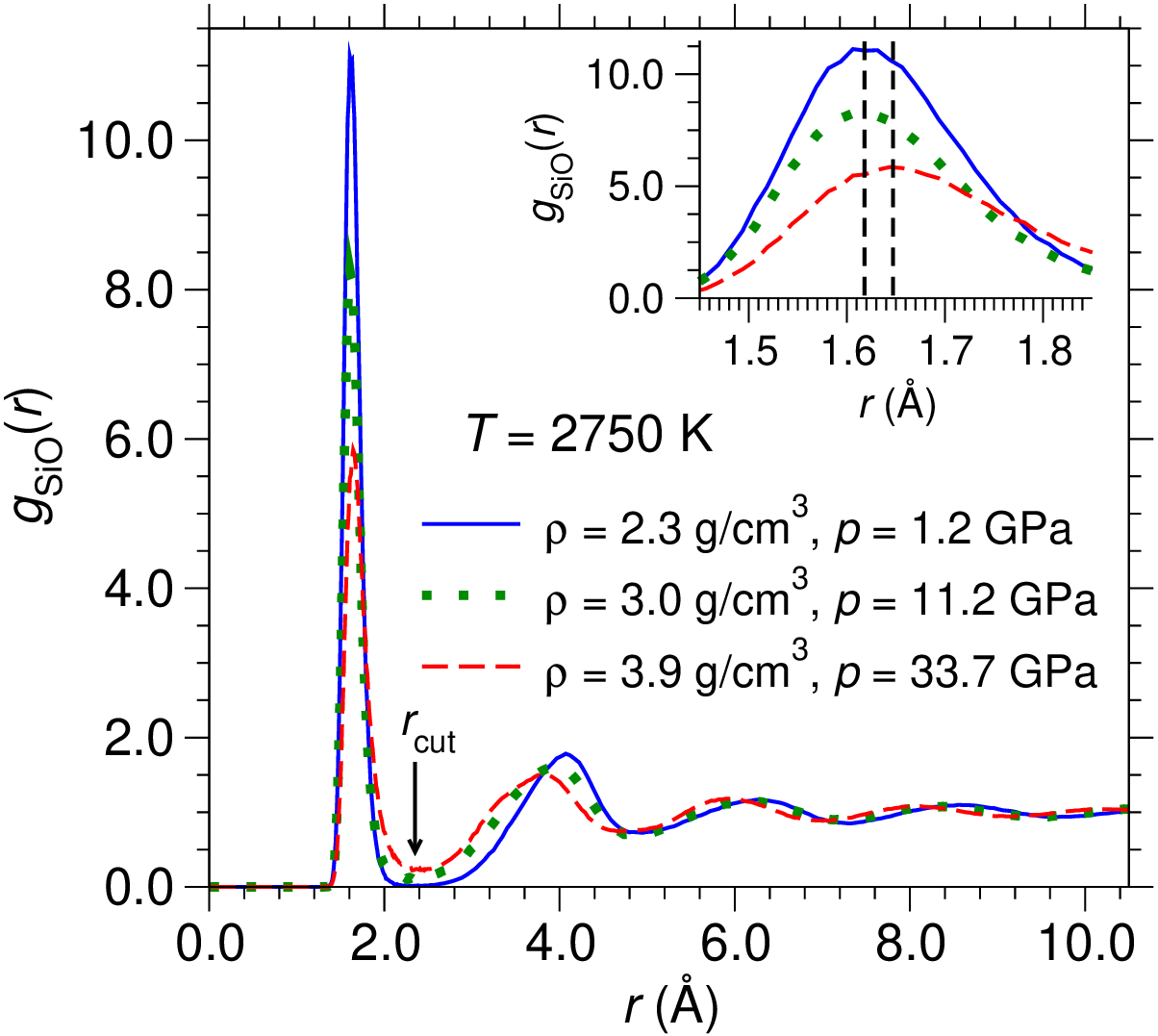}
\caption{Partial pair correlation function for the Si-O correlations,
$g_{\rm SiO}(r)$, for different densities at the temperature
$T=2750$\,K. The arrow indicates the cut-off $r_{\rm cut}=2.35$\,g/cm$^3$,
used to identify SiO bonds. The inset shows an enlargement of the
data. The two vertical dashed lines at 1.618\,\AA~and 1.647\,\AA~indicate
the location of the maxima for the densities $\rho=2.3$\,g/cm$^3$ and
$\rho=3.9$\,g/cm$^3$, respectively.
\label{fig2}
}
\end{center}
\end{figure}
\begin{figure}
\begin{center}

\vspace*{0.2cm}
\hspace*{1.9cm}
\includegraphics[width=0.4\textwidth]{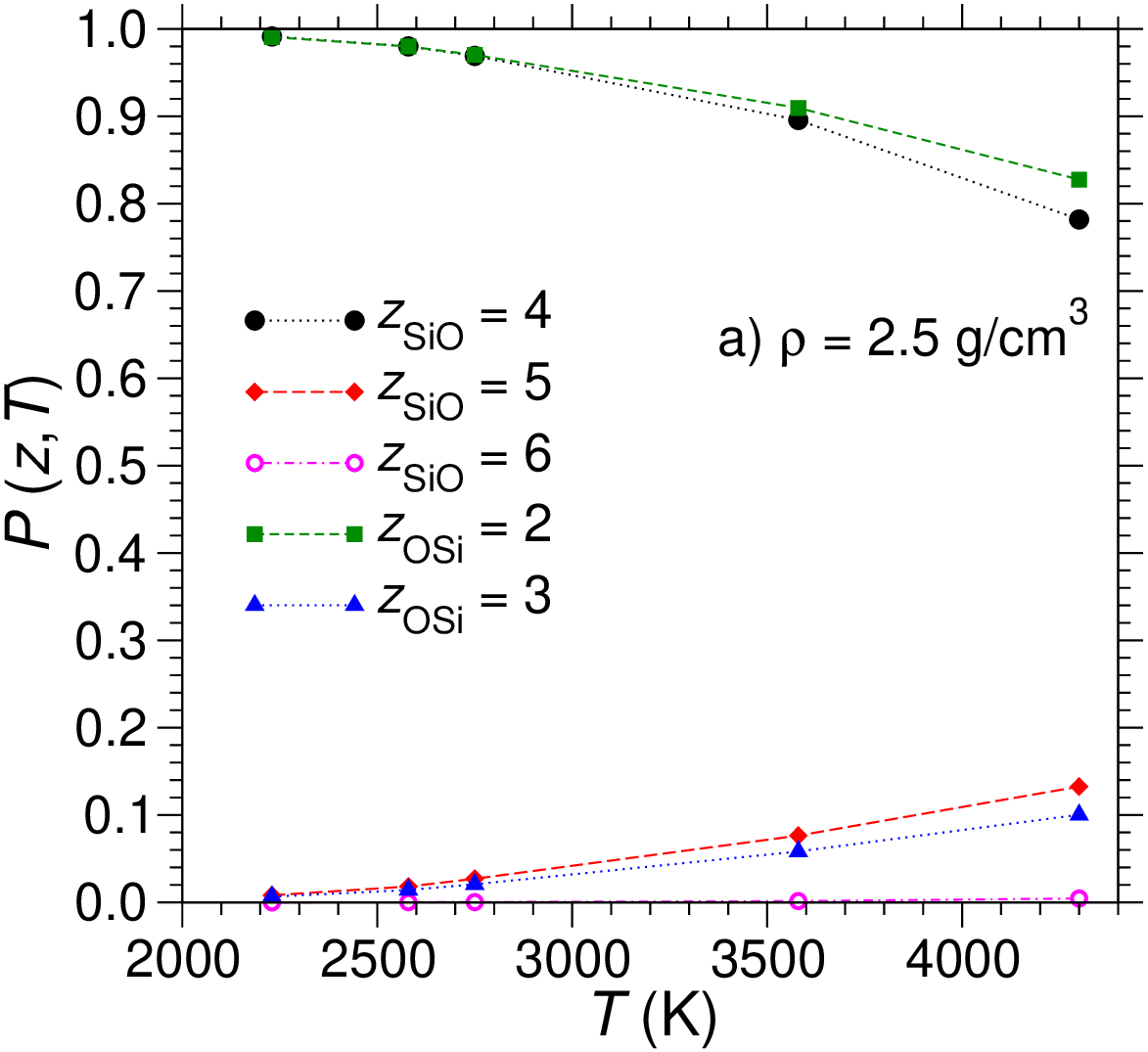}
\hspace*{0.2cm}
\includegraphics[width=0.4\textwidth]{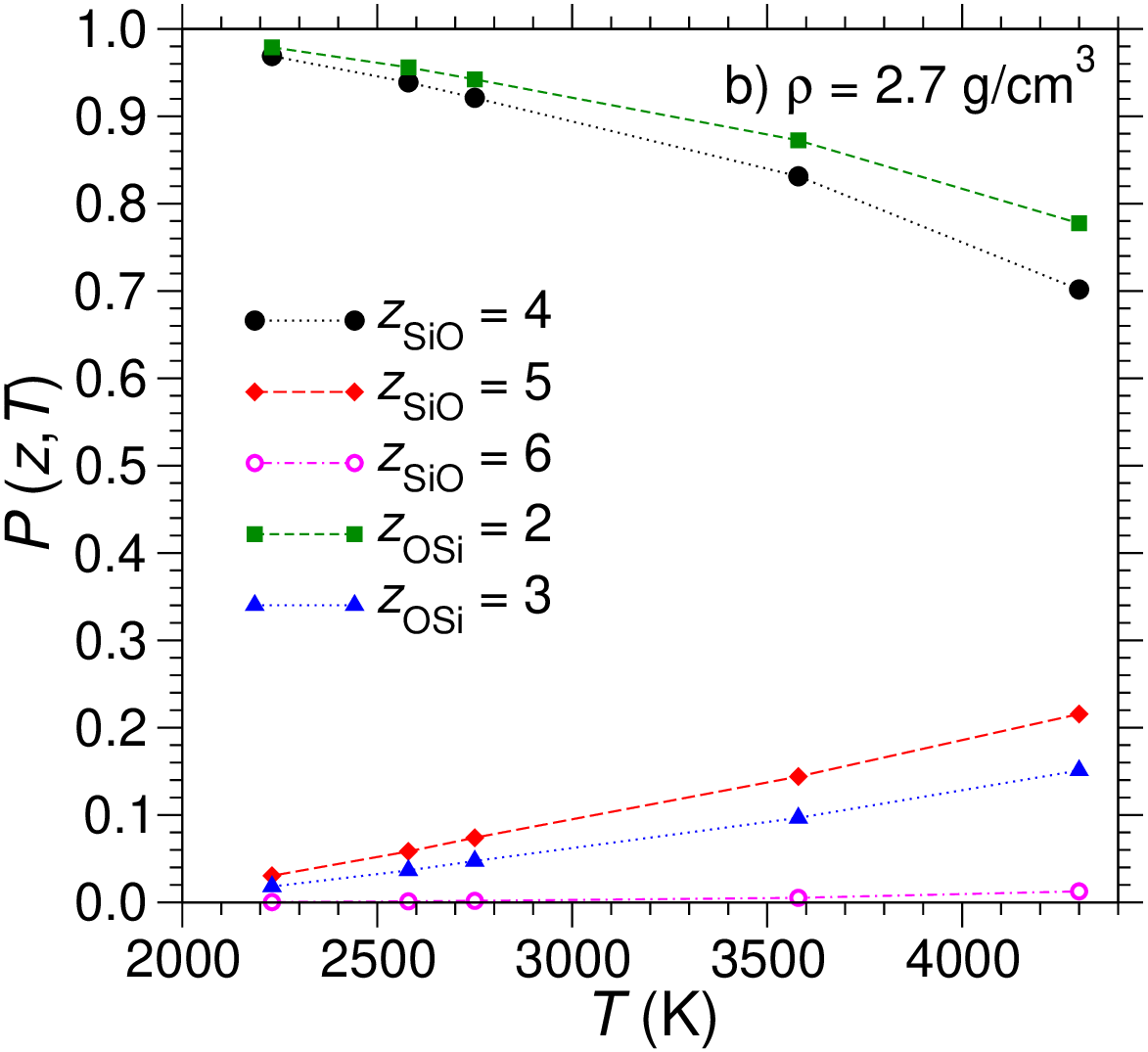}
\vspace*{0.5cm}

\hspace*{1.9cm}
\includegraphics[width=0.4\textwidth]{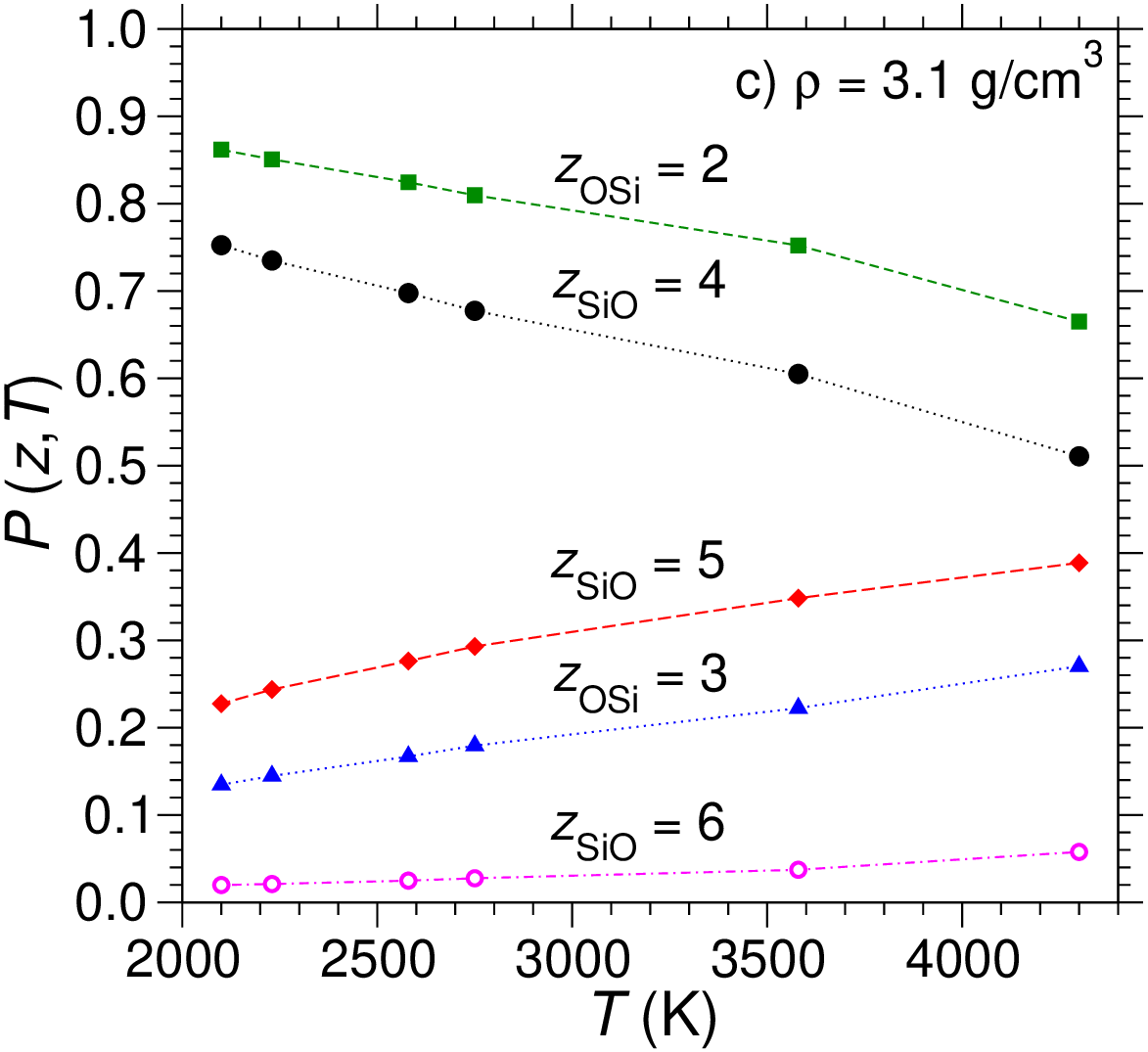}
\hspace*{0.2cm}
\includegraphics[width=0.4\textwidth]{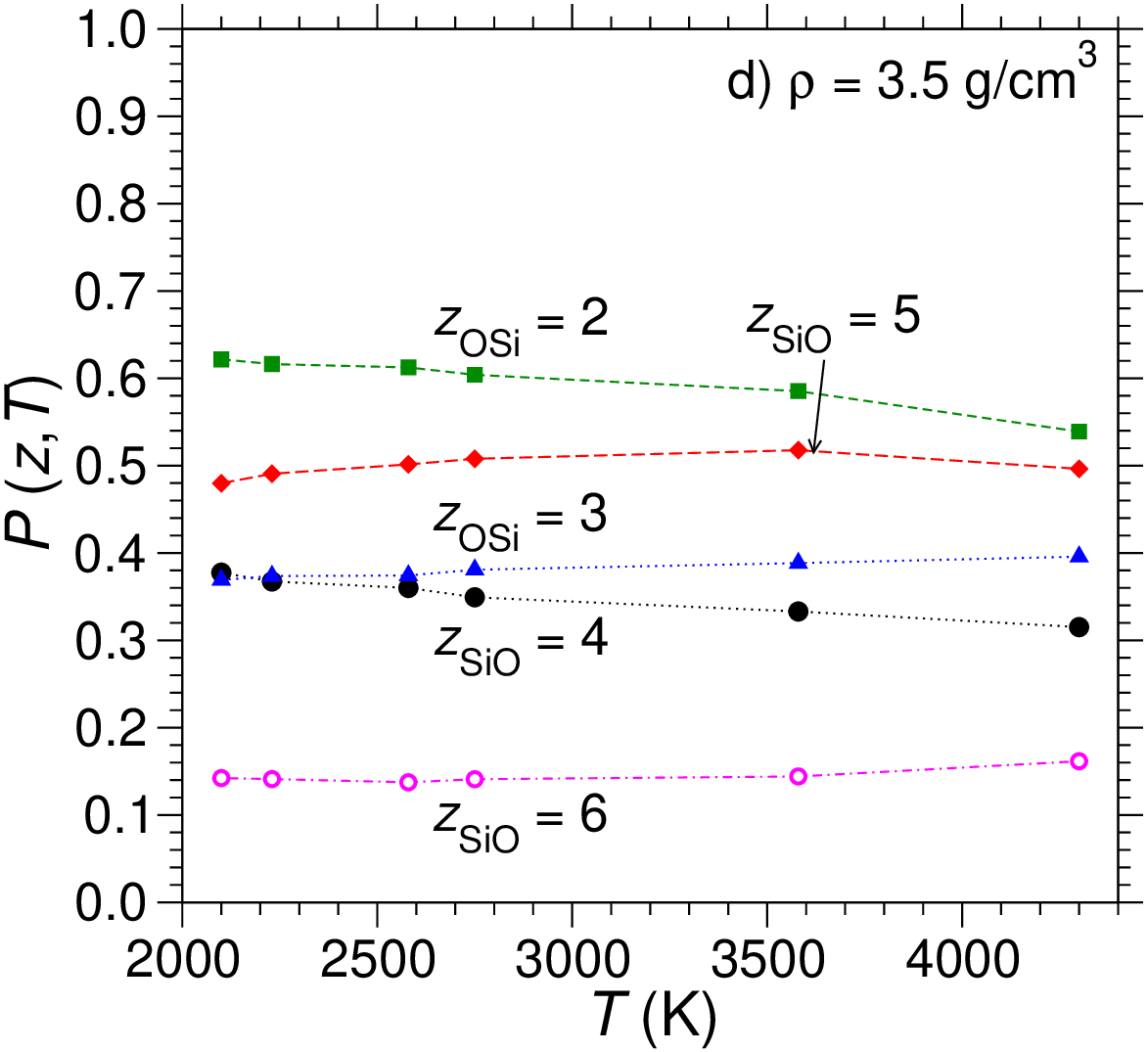}
\vspace*{0.5cm}

\hspace*{1.9cm}
\includegraphics[width=0.4\textwidth]{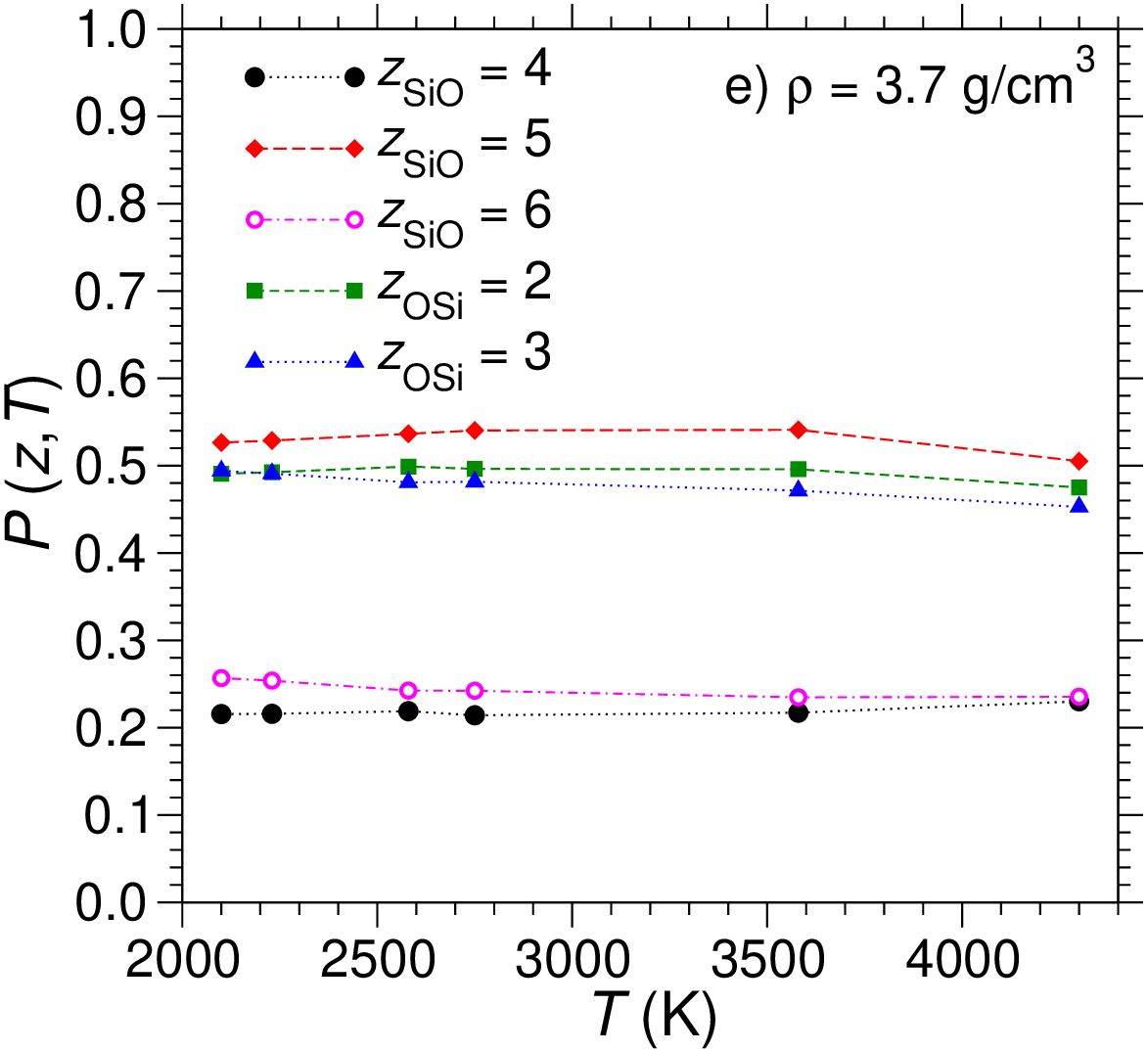}
\hspace*{0.2cm}
\includegraphics[width=0.4\textwidth]{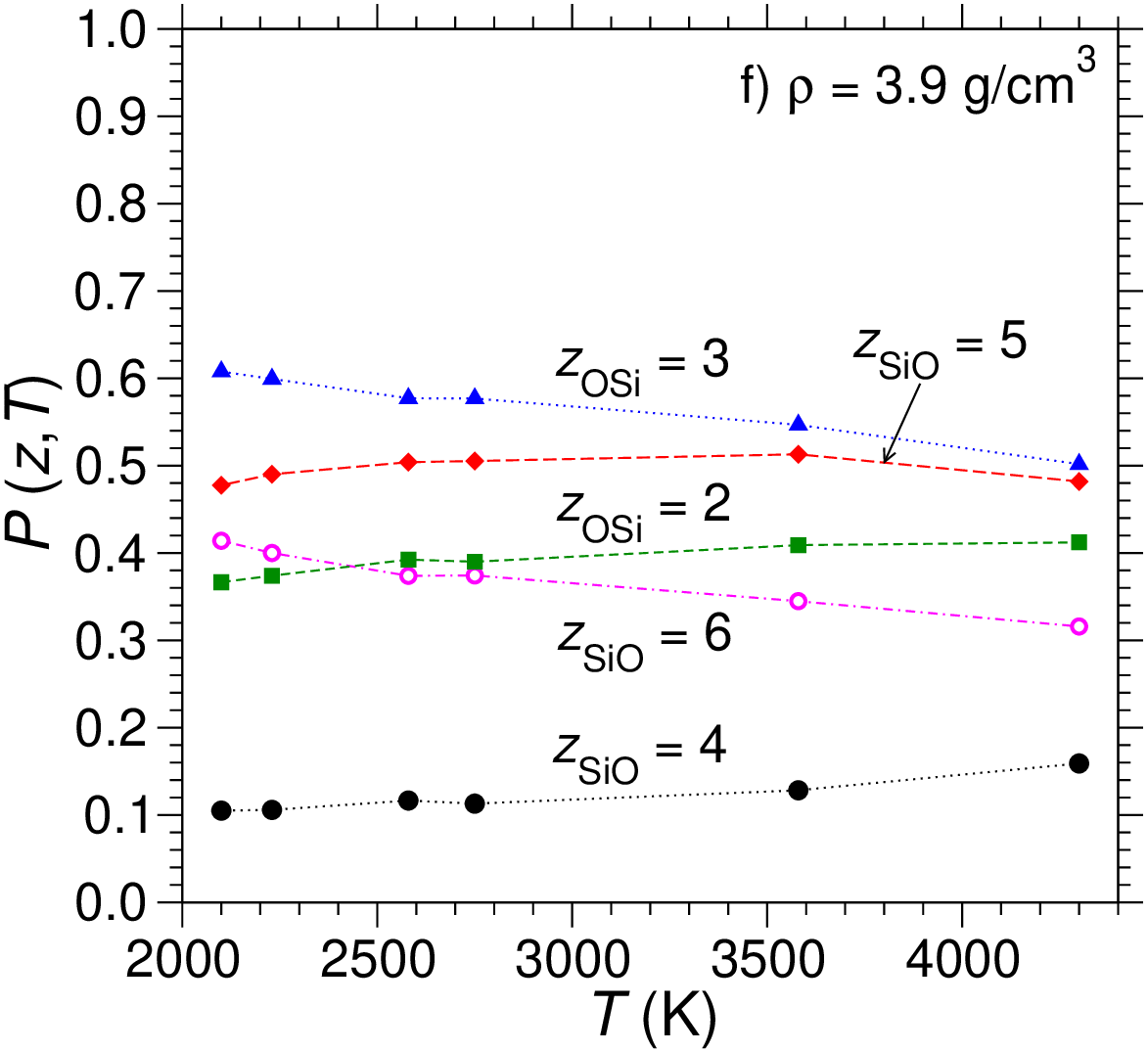}

\caption{Temperature dependence of different SiO and OSi coordination numbers $z$.
a) $\rho=2.5$\,g/cm$^3$, b) $\rho=2.7$\,g/cm$^3$, 
c) $\rho=3.1$\,g/cm$^3$, d) $\rho=3.5$\,g/cm$^3$,
e) $\rho=3.7$\,g/cm$^3$, f) $\rho=3.9$\,g/cm$^3$.
\label{fig3}
}
\end{center}
\end{figure}
\begin{figure}
\vspace*{0.26cm}
\begin{center}
\includegraphics[width=0.65\textwidth]{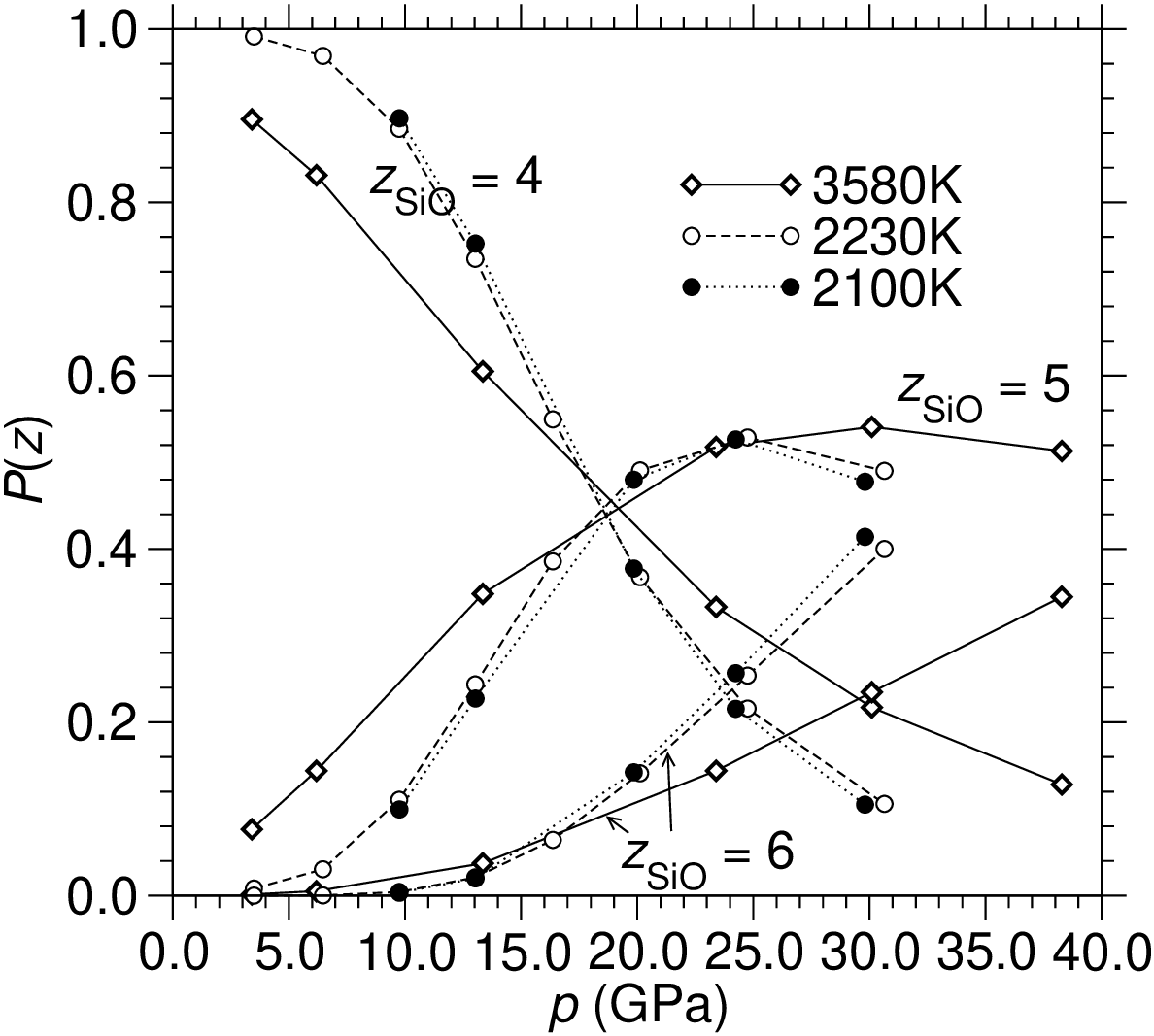}
\caption{Pressure dependence of the Si-O coordination at $T=3580$\,K, 2230\,K,
and 2100\,K.
\label{fig4}
}
\end{center}
\end{figure}
\begin{figure}
\begin{center}

\vspace*{0.2cm}
\hspace*{1.9cm}
\includegraphics[width=0.4\textwidth]{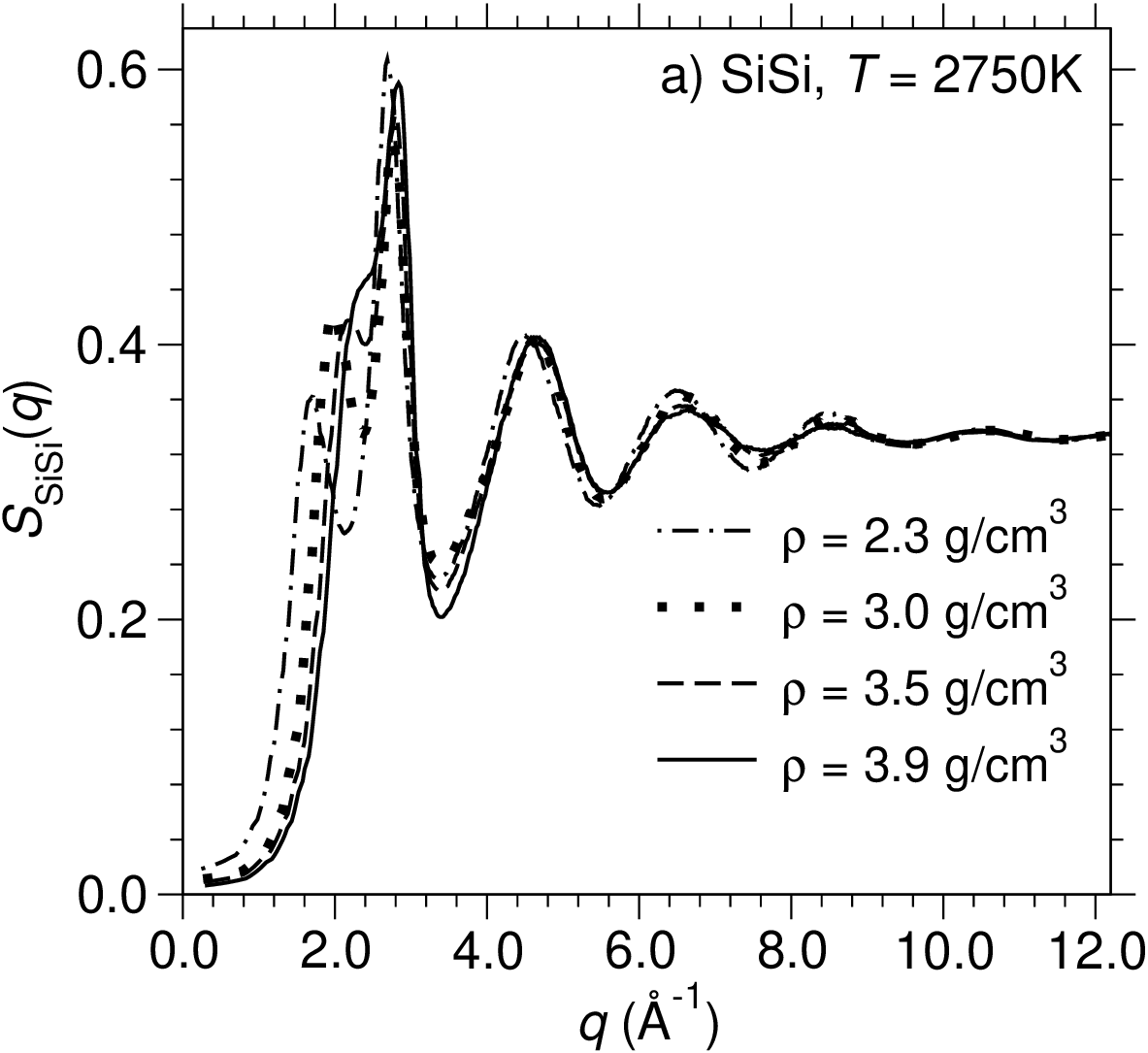}
\hspace*{0.2cm}
\includegraphics[width=0.4\textwidth]{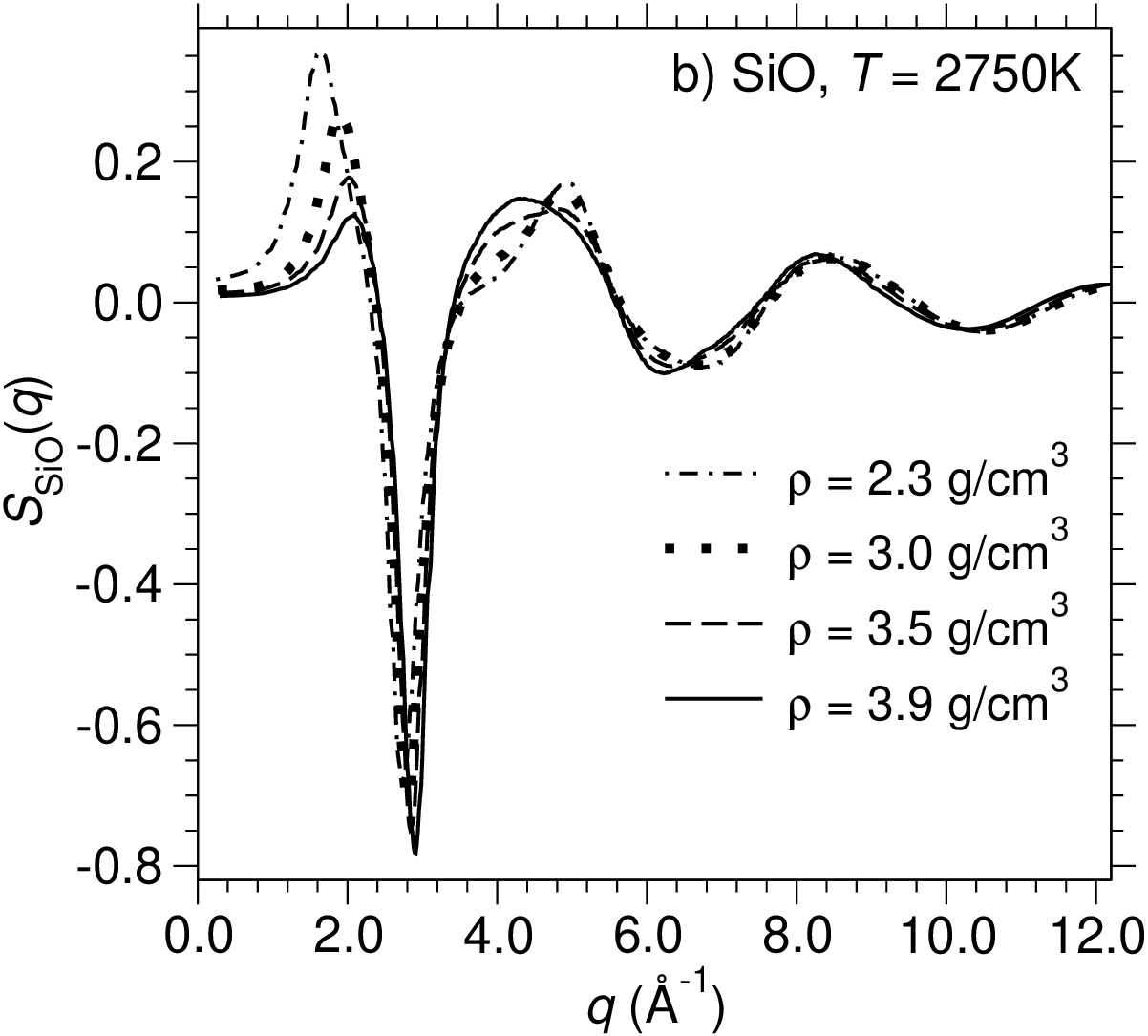}
\vspace*{0.5cm}

\hspace*{1.9cm}
\includegraphics[width=0.4\textwidth]{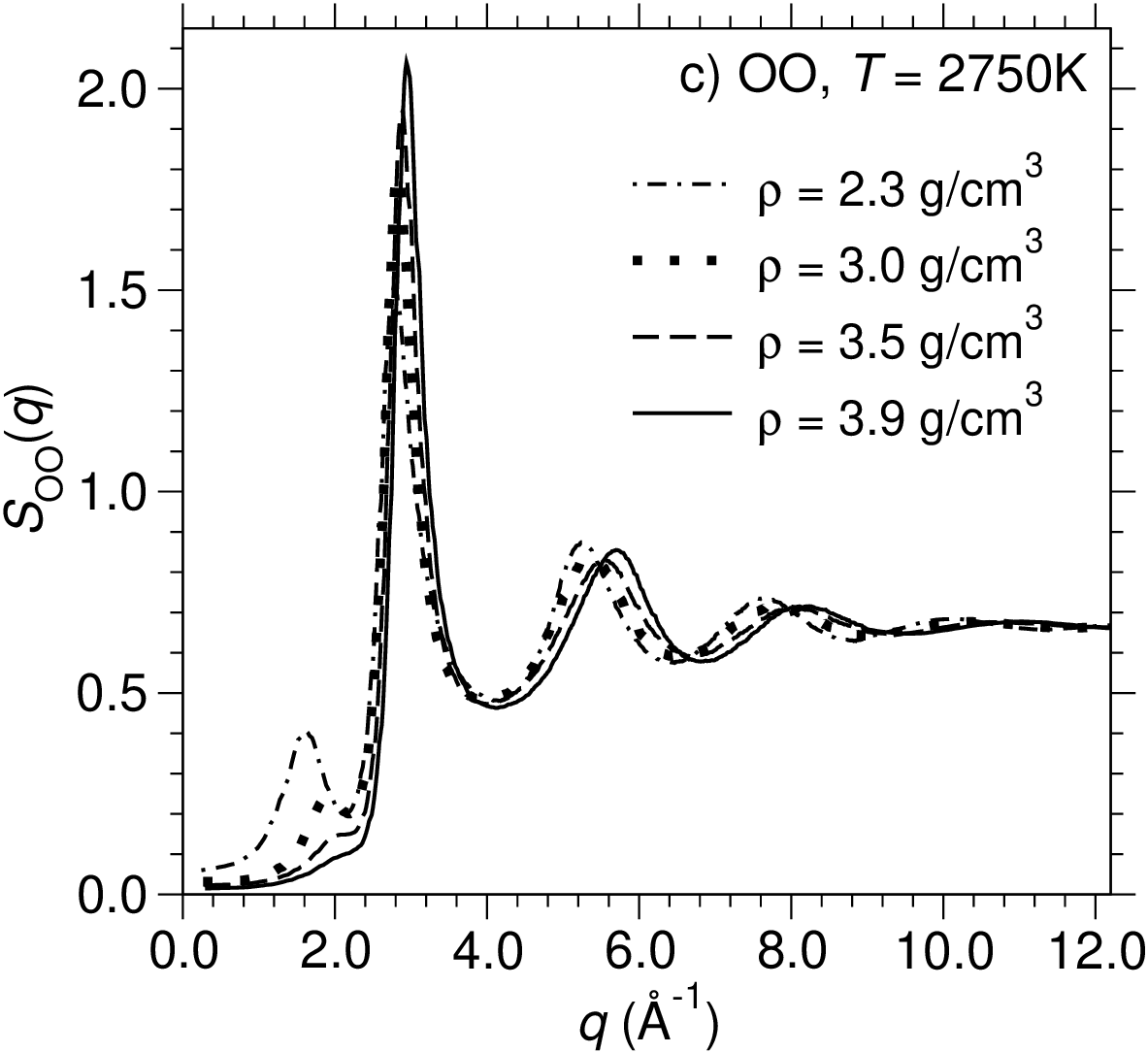}
\hspace*{0.2cm}
\includegraphics[width=0.4\textwidth]{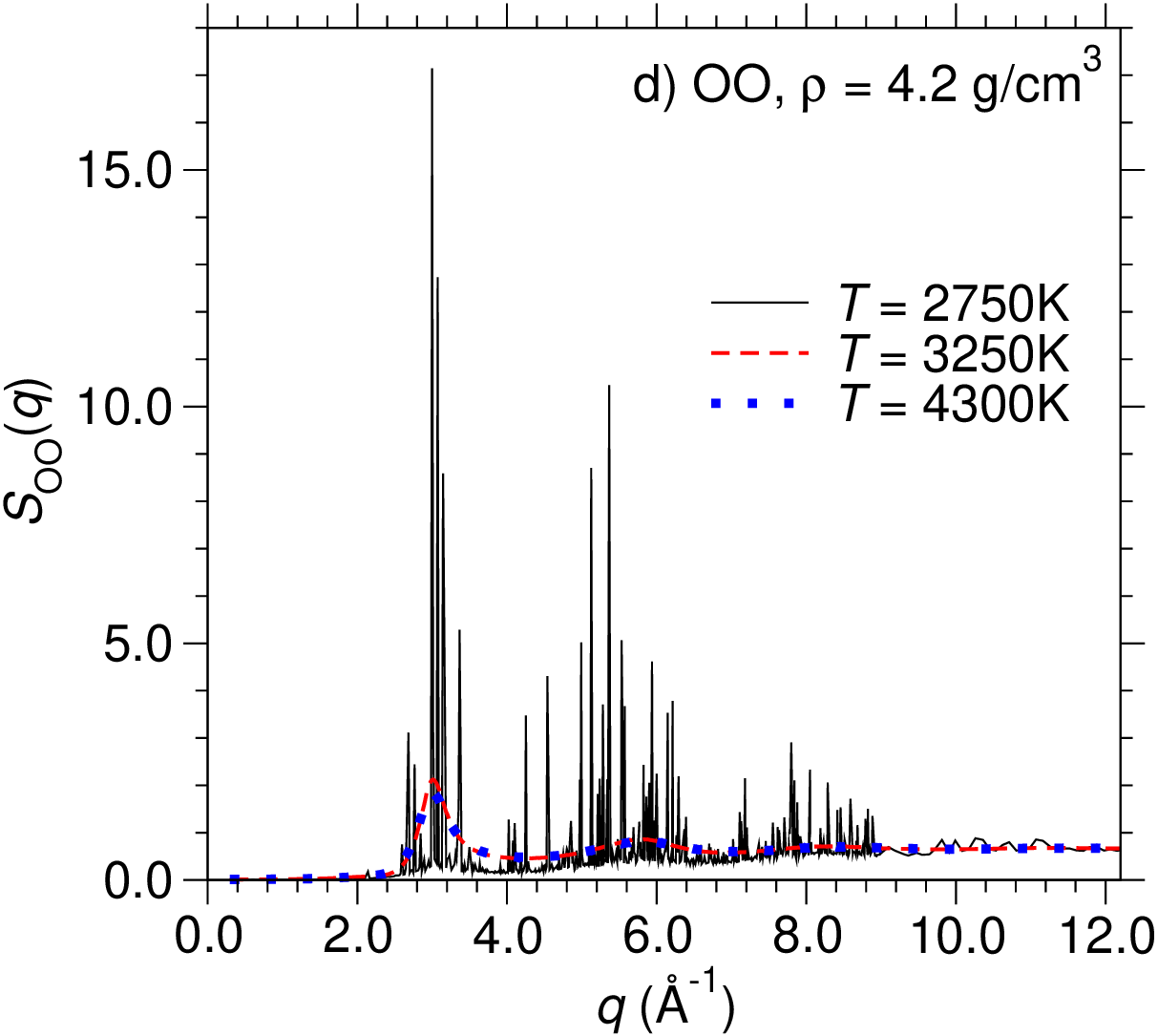}

\caption{Partial static structure factors $S_{\alpha\beta}(q)$ 
($\alpha, \beta = {\rm Si, O}$) for different densities at the temperature 
$T=2750$\,K. a) $S_{\rm SiSi}(q)$,  b) $S_{\rm SiO}(q)$,  c) $S_{\rm OO}(q)$.  
In panel d), $S_{\rm OO}(q)$ for $\rho=4.2$\,g/cm$^3$ at three different 
temperatures is shown, indicating that the system has crystallized at 
$T=2750$\,K. 
\label{fig5}
}
\end{center}
\end{figure}
\section{Results}
To analyze the change of the network structure of amorphous silica under
compression, we consider the temperature dependence of the Si-O
as well as the O-Si coordination.  To this end, coordination numbers
$z_{\rm SiO}$ and $z_{\rm OSi}$ are defined as the number of O (Si)
atoms surrounding a Si (O) atom within a distance $r\le r_{\rm cut}$.
For the cut-off radius we use the value $r_{\rm cut}=2.35$\,\AA~that
corresponds to the location of the first minimum of the partial pair
correlation function for the Si-O correlations, $g_{\rm SiO}(r)$
\cite{glassbook}.  As Fig.~\ref{fig2} shows, the location of this
minimum is indeed essentially independent of density.  However, one
can infer from the inset of Fig.~\ref{fig2} that the length of the SiO
bond changes slightly from about 1.62\,\AA~at 2.3\,g/cm$^3$ to about
1.65\,\AA~at 3.9\,g/cm$^3$. This finding is in very good agreement with
results from X-ray diffraction experiments \cite{meade92}.  Moreover,
the periodicity of the peaks in $g_{\rm SiO}(r)$ changes with density.
As we shall see below, this is due to the loss of tetrahedral order at
high densities, reflected, e.g., in the Fourier transform of $g_{\rm
SiO}(r)$, the partial structure factor $S_{\rm SiO}(q)$, by a shift of
the first sharp diffraction peak to higher wavenumbers $q$.

Figure \ref{fig3} shows the probability distributions $P(z)$ at different
densities for all the relevant coordination numbers $z_{\rm SiO}$ and
$z_{\rm OSi}$ as a function of temperature.  For the three densities
$\rho=2.5$, 2.7, and 3.1\,g/cm$^3$, most of the Si atoms exhibit the
tetrahedral four-fold coordination by O atoms and most of the O atoms
serve as bridging atoms between the tetrahedra ($z_{\rm OSi}=2$).  In this
case, the local defect structures that correspond to $z_{\rm SiO}=5, 6$ as
well as $z_{\rm OSi}=3$ become less frequent with decreasing temperature
such that one expects a very low probability of these structures
at low temperatures.  A different behavior is seen at the density
$\rho=3.5$\,g/cm$^3$.  Now, $z_{\rm SiO}=5$ has a higher probability
than the tetrahedral coordination $z_{\rm SiO}=4$ and one observes a
relatively high probability of $z_{\rm OSi}=3$ and a significant number
of six-fold coordinated Si atoms.  One can hardly extrapolate the curves
for 3.5\,g/cm$^3$ to lower temperatures since at this density all the
coordination numbers show a rather weak temperature dependence in the
considered temperature range. The same trend is also seen at the two high
densities, $\rho=3.7$\,g/cm$^3$ and $\rho=3.9$\,g/cm$^3$.  Whereas the
number of five-fold coordinated Si atoms keeps approximately constant
around 50\%, the coordination number $z_{\rm SiO}=6$ further increases,
whereas the tetrahedral coordination $z_{\rm SiO}=4$ becomes less and
less important with increasing density. At $\rho=3.9$\,g/cm$^3$, there
are more oxygen atoms that are three-fold coordinated by Si atoms than
bridging oxygens.

The SiO coordination as a function of pressure is shown in Fig.~\ref{fig4}
for $T=2100$\,K, 2230\,K and 3580\,K.  Although the qualitative behaviour
of $P(z_{\rm SiO})$ is similar at the different temperatures, there are
significant quantitative differences between the coordination numbers
at $T=3580$\,K and those at the two lower temperatures.  The maximum
of $P(z_{\rm SiO}=5)$ shifts from about 30\,GPa at 3580\,K to about
25\,GPa at the two lower temperatures. Moreover, at low temperatures,
$P(z_{\rm SiO}=4)$ decays more rapidly, accompanied by a faster increase
of $P(z_{\rm SiO}=6)$.  Note that in recent simulation studies using both
ab initio techniques and classical MD simulation \cite{trave02,tangney02},
the pressure dependence of the SiO coordination has been only discussed
for temperatures around 3500\,K. However, Fig.~\ref{fig4} demonstrates
that at least in the temperature range between 2000 to 4000\,K the
temperature dependence of $P(z_{\rm SiO})$ has to be taken into account
to allow for a quantitative comparison with experimental data, done at
much lower temperatures.

\begin{figure}

\begin{center}
\vspace*{0.3cm}
\includegraphics[width=0.6\textwidth]{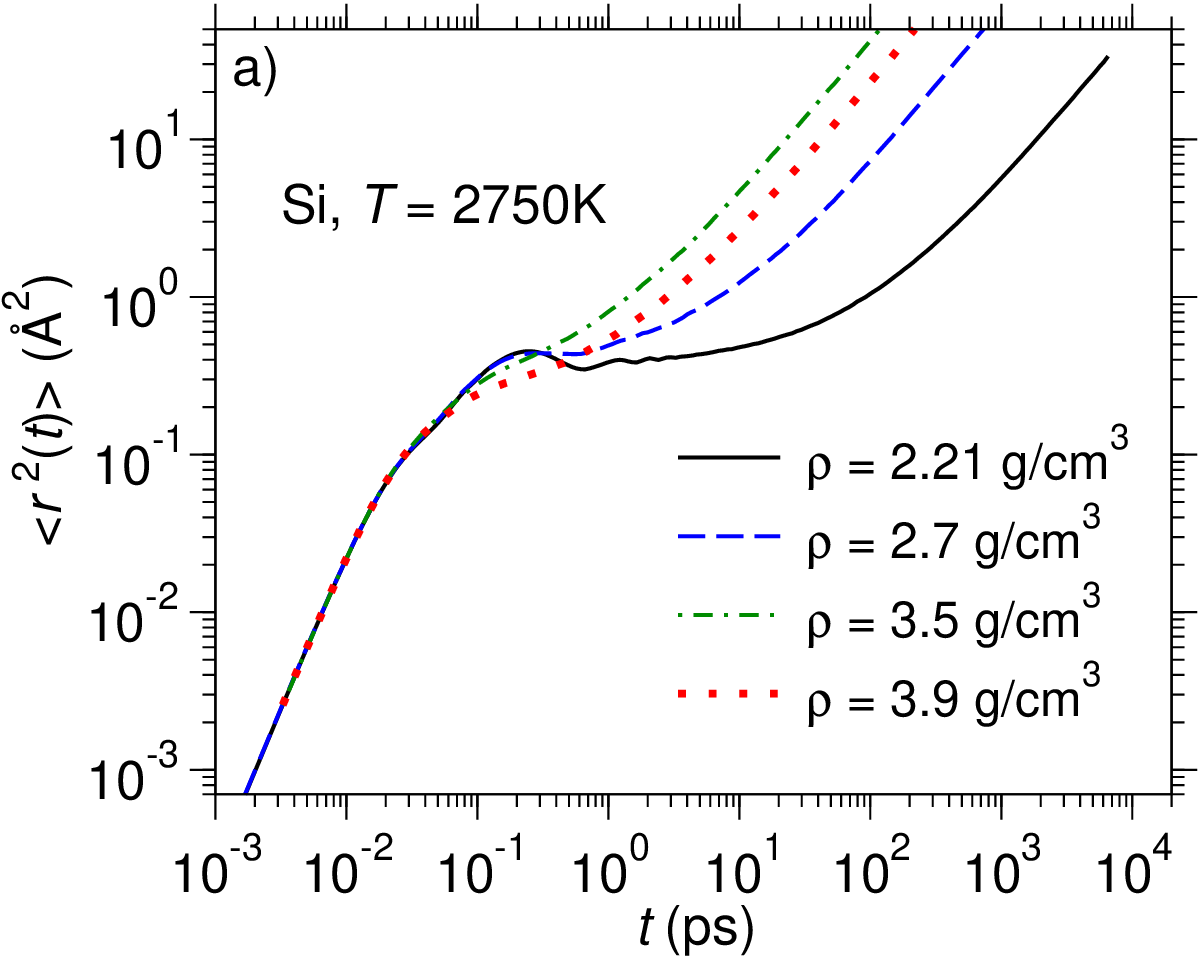}
\includegraphics[width=0.6\textwidth]{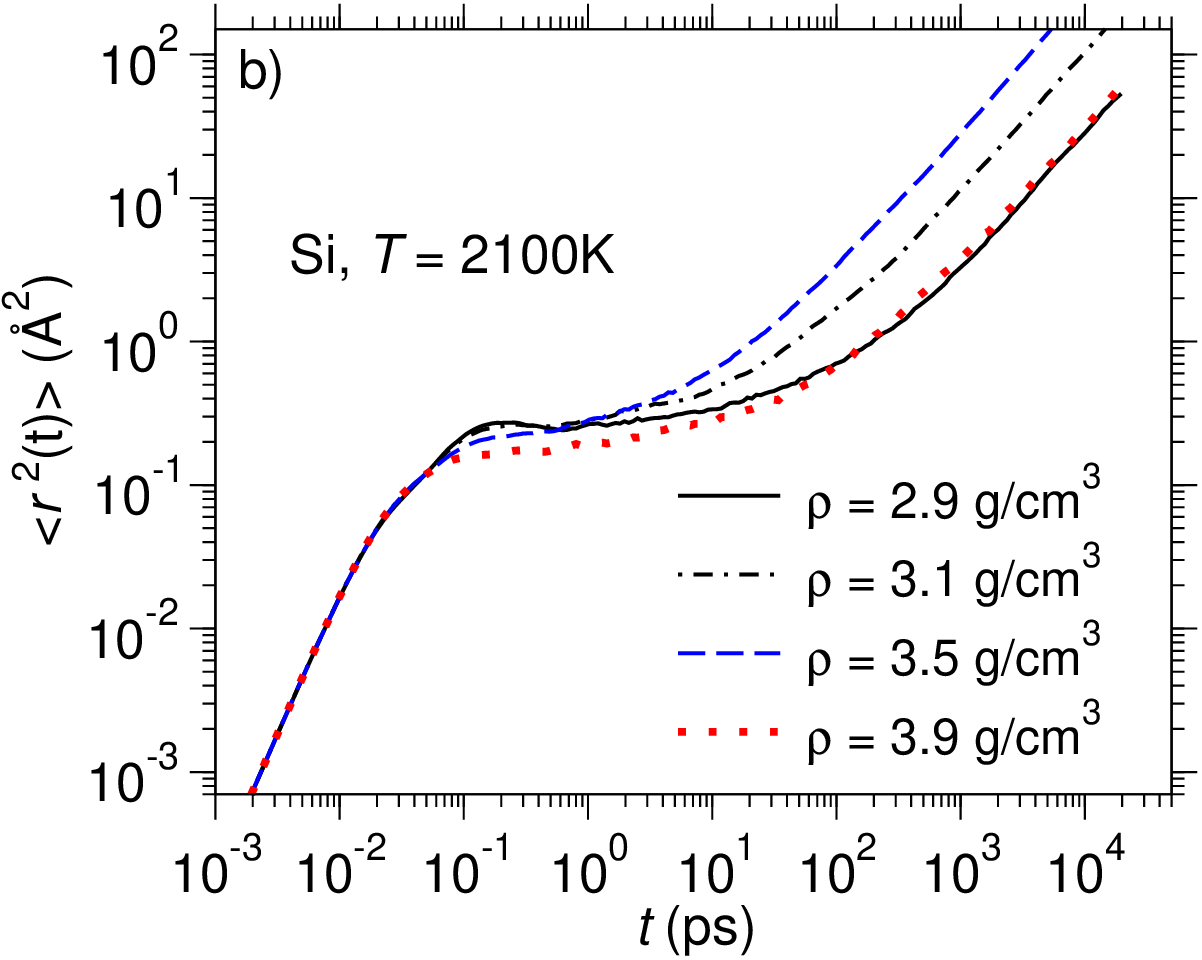}

\caption{Mean squared displacements for the silicon atoms for different densities
at a) $T=2750$\,K and b) $T=2100$\,K. 
\label{fig6}
}
\end{center}
\end{figure}
The inspection of the SiO and OSi coordination number distributions
indicates that significant changes occur in the network structure for
pressures above, say 10\,GPa. This finding is similar to what has been
observed experimentally for silica glasses. At pressures above about
20\,GPa a network with five- and six-fold coordinated Si atoms is formed,
similar to what has been recently found in other simulation studies
\cite{barrat97,trave02,tangney02}.  It remains an open question whether
in silica a first-order phase transition from a low density liquid to a
high density liquid exists.  Our data in Figs.~\ref{fig3} and \ref{fig4}
is also consistent with a gradual crossover from a low density phase
(with $z_{\rm SiO}=4$) to a high density phase (with $z_{\rm SiO}=6$),
occurring at low temperatures.

To investigate the changes in the network structure on intermediate
length scales, we now consider the partial static structure factors,
defined as \cite{glassbook}
\begin{equation}
S_{\alpha \beta}(q) = \frac{1}{N}
\sum_{k=1}^{N_{\alpha}} \sum_{l=1}^{N_{\beta}}
\left< \exp (i {\bf q} \cdot {\bf r}_{kl}) \right> \quad \quad 
\alpha, \beta = {\rm Si, O} ,
\end{equation}
with $N_{\alpha}$ the number of atoms of type $\alpha$, ${\bf q}$ the
wavevector, and ${\bf r}_{kl}={\bf r}_k - {\bf r}_l$ the distance vector
between particle $k$ and particle $l$.

Fig.~\ref{fig5} shows the three $S_{\alpha\beta}(q)$ at the temperature
$T=2750$\,K for different densities.  For the density $2.3$\,g/cm$^3$
the first peak [also called first sharp diffraction peak (FSDP)]
is located at a wavenumber of about 1.6-1.7\,\AA$^{-1}$.  This peak
is due to the tetrahedral order present in ``low-density'' silica.
Its location is related to the length scale of two connected tetrahedra
\cite{horbach99}. With increasing density, the first peak shifts to higher
values of $q$ and its amplitude decreases significantly.
The behavior of the FSDP reflects the change of the structure from an open
tetrahedral network at low densities to a more densely packed network at
high densities, provided by SiO$_5$ and SiO$_6$ units as well as oxygen
atoms that are three-fold coordinated by silicon atoms.  Note that the
partial structure factors $S_{\alpha\beta}(q)$ are the required input
for calculations in the framework of the mode-coupling theory of the
glass transition \cite{mct}, MCT, from which the dynamics of the melt
can be predicted. A MCT calculation based on the $S_{\alpha\beta}(q)$
of our simulation is presented elsewhere in this issue \cite{voigtmann07}.

Figure \ref{fig5}d shows $S_{\rm OO}(q)$ for the density
$\rho=4.2$\,g/cm$^3$ at three different temperatures. At this high
density, the system crystallizes at a temperature of $T=2750$\,K which
is reflected by the occurrence of Bragg peaks in $S_{\rm OO}(q)$. At
a temperature of $T=3250$\,K the system remains liquid; the structure
factor in this case reminds one to that of a simple liquid, showing no
sign of the FSDP feature.

\begin{figure}
\begin{center}

\includegraphics[width=0.6\textwidth]{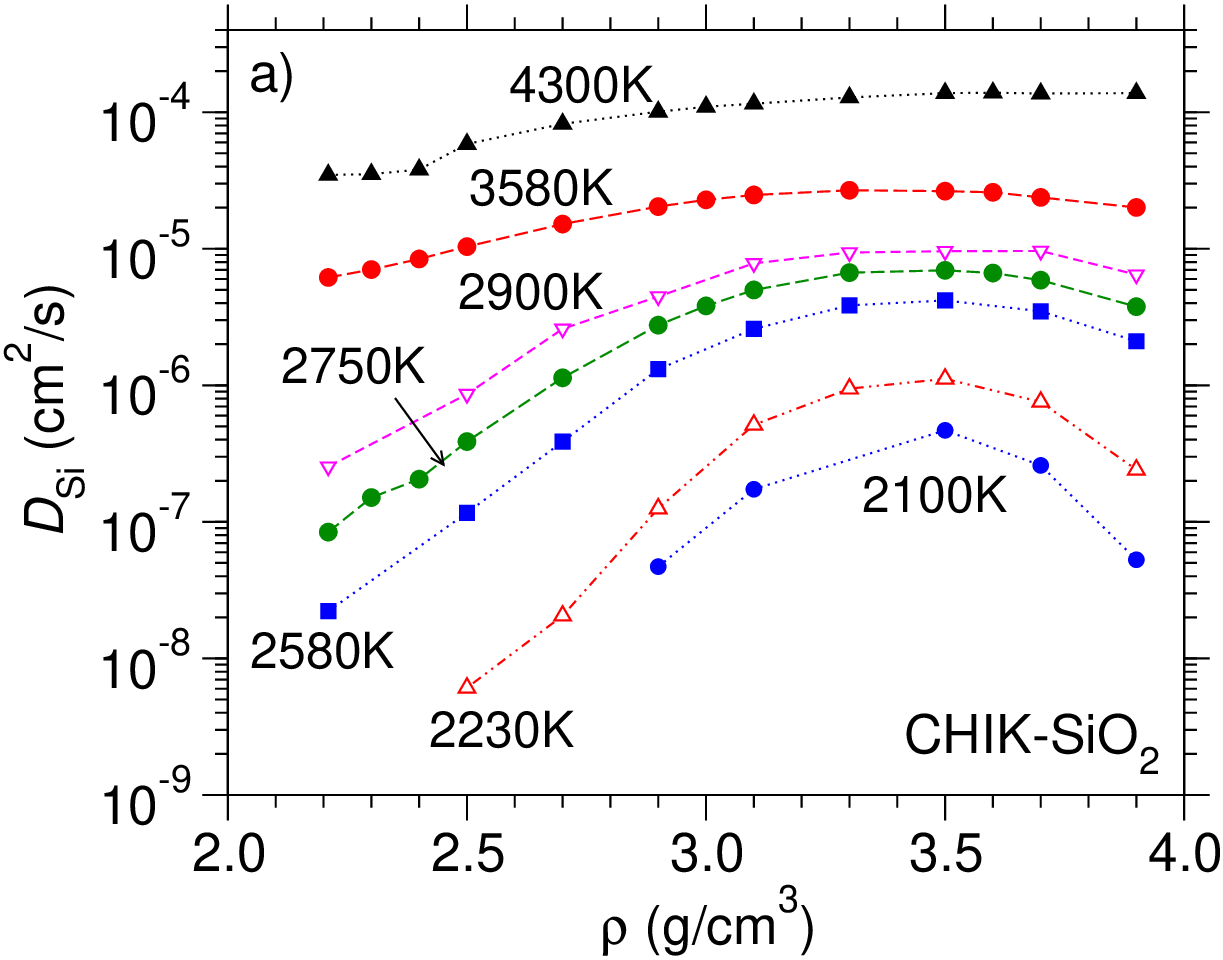}
\vspace*{0.3cm}

\includegraphics[width=0.6\textwidth]{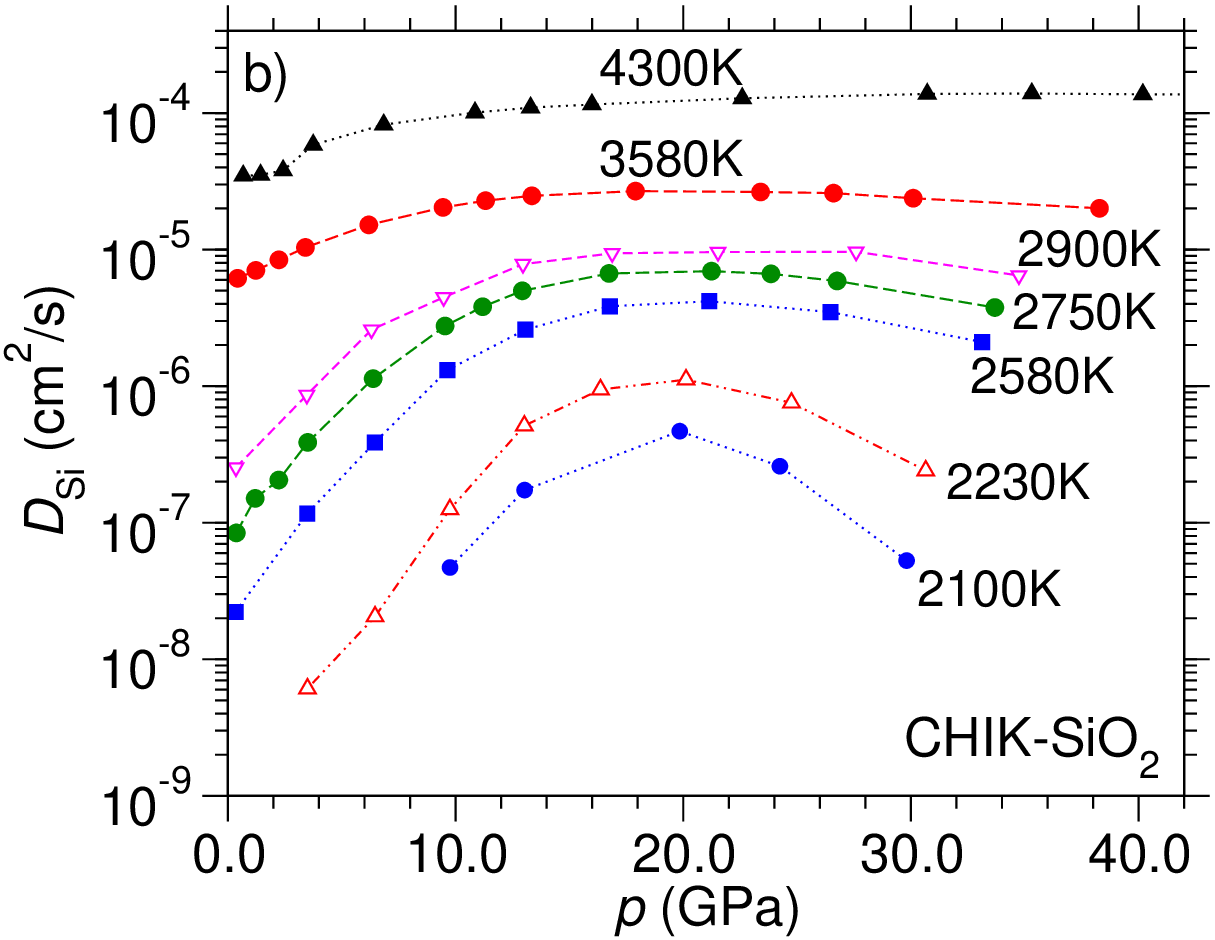}

\caption{Self-diffusion constant of silicon as a function of density 
along different isotherms as a function of density, a), and as a function 
of pressure, b). \label{fig7}
}
\end{center} 
\end{figure}
The structural changes in the network with increasing pressure are
accompanied by an anomalous behavior of relaxation processes. This
can be inferred from the mean squared displacements of a tagged
particle of type $\alpha={\rm Si,O}$, $\langle r^2_{\alpha}(t)\rangle=
(1/N_{\alpha}) \sum_{k=1}^{N_{\alpha}} \langle \left({\bf r}_k(t) - {\bf
r}_k(0) \right)^2 \rangle$. In Fig.~\ref{fig6} only $\langle r^2_{\rm
Si}(t)\rangle$ is shown, since the mean squared displacement for oxygen
exhibits a similar behaviour. At $T=2750$\,K, several time regimes can be
clearly distinguished at the low density $\rho=2.21$\,g/cm$^3$, consisting
of a ballistic regime at very short times ($\langle r^2(t) \rangle \propto
t^2$), an intermediate regime with a maximum 0.2\,ps, a plateau-like
region and finally a linear regime, $\langle r^2(t) \rangle \propto t^2$,
indicating diffusive behaviour.  Note that the maximum at 0.2\,ps is both
related to the so-called boson peak and to finite size effects present in
the dynamics of silica (for details see Refs.~\cite{horbach96,horbach01}).
That this feature is not seen at high density might be related to the
changes occurring in the network structure.  From $\rho=2.21$\,g/cm$^3$
to $\rho=3.5$\,g/cm$^3$, an acceleration of the diffusion dynamics can be
inferred from Fig.~\ref{fig6}a, followed by the opposite behaviour for a
further decrease of the density.  The same behaviour as for $T=2750$\,K is
also seen for the lower temperature, $T=2100$\,K, although the effect of
a maximal speed of diffusion is now more pronounced (Fig.~\ref{fig6}b).
One can also infer from Fig.~\ref{fig6}b that the particles become
more localized with increasing density, indicated by a decrease of the
plateau's height with increasing density.

The self-diffusion constants can be easily determined from the mean
squared displacements using the Einstein relation, $D_{\alpha}= \lim_{t\to
\infty} \langle r^2_{\alpha}(t) \rangle/(6t)$.  In Fig.~\ref{fig7},
the self-diffusion constant for silicon, $D_{\rm Si}$, along different
isotherms is displayed as a function of density and pressure.
Note that $D_{\rm O}$, albeit slightly larger, exhibits the same
behaviour.  Along the different isotherms a maximum is found around
about 3.5\,g/cm$^3$, corresponding to a pressure of about 20\,GPa
(see Fig.~\ref{fig7}b). Thereby, the diffusivity maximum becomes more
pronounced with decreasing temperature.  The behaviour of $D_{\rm Si}$
is closely related to that of the coordination numbers. Above about
$p=20$\,GPa, the location of the diffusivity maximum, the number of
five-fold coordinated Si atoms roughly shows a saturation, while
the number of six-fold coordinated Si atoms further increases with
increasing pressure.  It seems that a SiO network with many five-fold
coordinated Si atoms tends to have a faster diffusion dynamics than
one with many six-fold coordinated Si atoms. Therefore, for pressures
up to about 20\,GPa the network dynamics is dominated by an increasing
number of five-fold coordinated Si atoms, leading to an acceleration
of diffusion. However, at pressures above about 20\,GPa, the network
dynamics is dominated by an increasing number of SiO$_6$ units, leading
to a decrease of the diffusion constants with pressure. Thus, the
diffusivity maximum is due to an interplay between five- and six-fold
coordinated Si atoms.

\section{Conclusions}
Results of an extensive MD simulation have been presented to study
structural and dynamic properties of liquid silica under (high) pressure.
As an interaction potential a simple pair potential, the so-called
CHIK potential, was used that leads to a reliable description of the
equation of state. We have shown that the change of the network structure
with increasing pressure is related to an anomalous diffusion dynamics,
characterized by a diffusivity maximum around 20\,GPa.  Elsewhere in this
issue \cite{voigtmann07}, a calculation in the framework of mode coupling
theory based on the structural input of our simulation is presented that
explicitly shows that the changes in the network structure are intimately
related to the anomalous transport in amorphous silica.  It is still an
open question whether, at low temperatures, there is a first-order phase
transition from a low-density to a high-density transition in silica. This
question will be addressed in future studies with the CHIK potential.

\ack{We thank Schott Glas for financial support. Computing time on the JUMP
at the NIC J\"ulich is gratefully acknowledged.}
%

\section*{References}


\begin{thebibliography}{99}
%

%
\bibitem{grimsditch84}
Grimsditch M
1984 {\it Phys. Rev. Lett.} {\bf 52} 2379
%
\bibitem{hemley86}
Hemley R J, Mao H K, Bell P M and Mysen B O
1986 {\it Phys. Rev. Lett.} {\bf 57} 747
%
\bibitem{williams88}
Williams Q and Jeanloz R 1988
{\it Science} {\bf 239} 902
%
\bibitem{meade92}
Meade C, Hemley R J and Mao H K
1992 {\it Phys. Rev. Lett.} {\bf 69} 1387
%
\bibitem{zha94}
Zha C, Hemley R J, Mao H, Duffy T S and Meade C
1994 {\it Phys. Rev. B} {\bf 50} 13105 
%
\bibitem{brazhkin03}
Brazhkin V V and Lyapin A G
2003 {\it J. Phys.: Condens. Matter} {\bf 15} 6059
%
\bibitem{lin07}
Lin J-F, Fukui H, Prendergast D, Okuchi T, Cai Y Q, Hiraoka N, Yoo C-S, Trave A,
Eng P, Hu M Y and Chow P,
2007 {\it Phys. Rev. B} {\bf 75} 012201
%
\bibitem{bks90}
van Beest B W H, Kramer G J and van Santen R A
1990 {\it Phys. Rev. Lett.} {\bf 64} 1955
%
\bibitem{tse92}
Tse J S, Klug D D and Le Page Y
1992 {\it Phys. Rev. B} {\bf 46} 5933
%
\bibitem{barrat97}
Barrat J-L, Badro J and Gillet Ph
1997 {\it J. Comp. Simul.} {\bf 20} 17
%
\bibitem{hemley00}
Hemley R J, Badro J and Teter D M 2000
in {\it Physics Meets Mineralogy - Condensed Matter Physics in Geosciences}
ed Aoki H, Syono Y and Hemley R J (Cambridge: Cambridge University Press) 
p 173
%
\bibitem{horbach96}
Horbach J, Kob W, Binder K and Angell C A 1996
{\it Phys. Rev. E} {\bf 54} R5897
%
\bibitem{kushiro78}
Kushiro I 1978 {\it Earth Planet. Sc. Lett.} {\bf 41} 87
%
\bibitem{rubie93}
Rubie D C, Ross C R, Carroll M R and Elphick S C 1993
{\it Am. Mineral.} {\bf 78} 574
%
\bibitem{poe97}
Poe B T, McMillan P F, Rubie D C, Chakraborty S, Yarger J and Diefenbacher J
1997 {\it Science} {\bf 276} 1245
%
\bibitem{tinker03}
Tinker D, Lesher C E and Hutcheon I D 2003 
{\it Geochim. Cosmochim. Acta} {\bf 67} 133
%
\bibitem{angell85}
Angell C A 1985
in {\it Relaxation in Complex Systems} 
ed Ngai K and Wright G B (Springfield: US Dept. of Commerce) 
%
\bibitem{glassbook}
Binder K and Kob W 2005
{\it Glassy Materials and Disordered Solids: An Introduction
to Their Statistical Mechanics} 
(Singapure: World Scientific)
%
\bibitem{angell92}
Angell C A, Cheeseman P A and Tammadon S 1992
{\it Science} {\bf 218} 885
%
\bibitem{saika00}
Saika-Voivod I, Sciortino F and Poole P H 2000
{\it Phys. Rev. E} {\bf 63} 011202
%
\bibitem{shell02}
Scott Shell M, Debenedetti P G and Panagiotopoulos A Z
2002 {\it Phys. Rev. E} {\bf 66} 011202
%
\bibitem{sharma06}
Sharma R, Mudi A and Chakravarty C,
2006 {\it J. Chem. Phys.} {\bf 125} 044705
%
\bibitem{legrand07}
Le Grand A, Dreyfus C, Bousquet C and Pick R M,
2007 {\it Phys. Rev. E} {\bf 75} 061203
%
\bibitem{kohn96}
Kohn W 1996 in
{\it Monte Carlo and Molecular Dynamics of Condensed Matter 
Systems} ed Binder K and Ciccotti (Bologna: Societ\`a Italiana di
Fisica) p 561; Car R {ibid.} p 601
%
\bibitem{trave02}
Trave A, Tangney P, Scandolo S, Pasquarello A and Car R 
2002 {\it Phys. Rev. Lett.} {\bf 89} 245504
%
\bibitem{tangney02}
Tangney P and Scandolo S 2002
{\it J. Chem. Phys.} {\bf 117} 8898
%
\bibitem{herzbach05}
Herzbach D, Binder K and M\"user M H 
2005 {\it J. Chem. Phys.} {\bf 123} 124711
%
\bibitem{carre07_1}
Carr\'e A, Horbach J, Ispas S and Kob W 2008
{\it Europhys. Lett.} {\bf 82} 17001
%
\bibitem{vollmayr96}
Vollmayr K, Kob W and Binder K 1996
{\it Phys. Rev. B} {\bf 54} 15808
%
\bibitem{horbach99}
Horbach J and Kob W 1999
{\it Phys. Rev. B} {\bf 60} 3169
%
\bibitem{horbach01}
Horbach J, Kob W and Binder K 2001
{\it Eur. Phys. J. B} {\bf 19} 531
%
\bibitem{allen}
Allen M and Tildesley D 1987
{\it Compute Simulation of Liquids}
(Oxford: Oxford University Press)
%
\bibitem{tsiok98}
Tsiok O B, Brazhkin V V, Lyapin A G and Khvostantsev L G 1998
{\it Phys. Rev. Lett.} {\bf 80} 999
%
\bibitem{mct}
G\"otze W 1999 
{\it J. Phys.: Condens. Matter} {\bf 11} A1
%
\bibitem{voigtmann07}
Voigtmann Th and Horbach J 2008
{\it J. Phys.: Condens. Matter} this issue
%

%
\end{thebibliography}
\end{document}